\documentclass[conference]{IEEEtran}
\IEEEoverridecommandlockouts

\usepackage{cite}
\usepackage{amsmath,amssymb,amsfonts}
\usepackage{algorithmic}
\usepackage{graphicx}
\usepackage{textcomp}
\usepackage{xcolor}
\usepackage{graphicx}
\usepackage{amsmath}
\usepackage{amssymb}
\usepackage{algorithm}
\usepackage{algorithmic}
\usepackage[utf8]{inputenc}
\usepackage{authblk}
\usepackage{amsmath}
\usepackage{amssymb}
\usepackage{setspace}  
\usepackage{float} 
\def\BibTeX{{\rm B\kern-.05em{\sc i\kern-.025em b}\kern-.08em
    T\kern-.1667em\lower.7ex\hbox{E}\kern-.125emX}}

\title{EnStack: An Ensemble Stacking Framework of Large Language Models for Enhanced Vulnerability Detection in Source Code}

\author[1]{Shahriyar Zaman Ridoy}
\author[2]{Md. Shazzad Hossain Shaon}
\author[3,\textsuperscript{\textsection}]{Alfredo Cuzzocrea}
\author[4,*]{Mst Shapna Akter}

\affil[1]{\small Department of Electrical and Computer Engineering, North South University, Dhaka, Bangladesh}
\affil[2]{\small Department of Computer Science and Engineering, Oakland
University, Rochester, MI 48309, USA}
\affil[3]{\small iDEA Lab, University of Calabria, Rende, Italy, and the Department of Computer Science, University of Paris City, Paris, France}
\affil[4]{\small Department of Computer Science, Oakland University, MI, USA}

\affil[ ]{\small shahriyar.ridoy@northsouth.edu, shaon@oakland.edu, alfredo.cuzzocrea@unical.it, akter@oakland.edu}

\date{}  

\begin{document}

\maketitle
\begingroup\renewcommand\thefootnote{\textsection}

\footnotetext{This research has been made in the context of the Excellence Chair in Big Data Management and Analytics at University of Paris City, Paris, France.}
\begin{abstract}
Automated detection of software vulnerabilities is
critical for enhancing security, yet existing methods often strug-
gle with the complexity and diversity of modern codebases. In this paper, we introduce EnStack, a novel ensemble stacking framework that enhances vulnerability detection using natural language processing (NLP) techniques. Our approach synergizes multiple pre-trained large language models (LLMs) specialized in code understanding—CodeBERT for semantic analysis, GraphCodeBERT for structural representation, and UniXcoder for cross-modal capabilities. By fine-tuning these models on the Draper VDISC dataset and integrating their outputs through meta-classifiers such as Logistic Regression, Support Vector Machines (SVM), Random Forest, and XGBoost, EnStack effectively captures intricate code patterns and vulnerabilities that individual models may overlook. The meta-classifiers consolidate the strengths of each LLM, resulting in a comprehensive model that excels in detecting subtle and complex vulnerabilities across diverse programming contexts. Experimental results demonstrate that EnStack significantly outperforms existing methods, achieving notable improvements in accuracy, precision, recall and F1-score. This work highlights the potential of ensemble LLM approaches in code analysis tasks and offers valuable insights into applying NLP techniques for advancing automated vulnerability detection.
\end{abstract}


\section{Introduction}

The increasing prevalence of software vulnerabilities in today's rapidly advancing software development landscape poses significant security threats to individuals, organizations, and governments alike~\cite{perwej2021systematic}. These vulnerabilities, often resulting from flaws in code implementation or design, can lead to serious consequences such as data breaches, system failures, and substantial financial losses. Traditional vulnerability detection methods, including manual code inspections and static analysis tools, are becoming less effective in managing the complexity and scale of modern software systems~\cite{ghaffarian2017software}. This situation underscores the urgent need for advanced techniques capable of efficiently identifying vulnerabilities within source code.

\begin{figure*}
    \centering
    \includegraphics[width=1\linewidth]{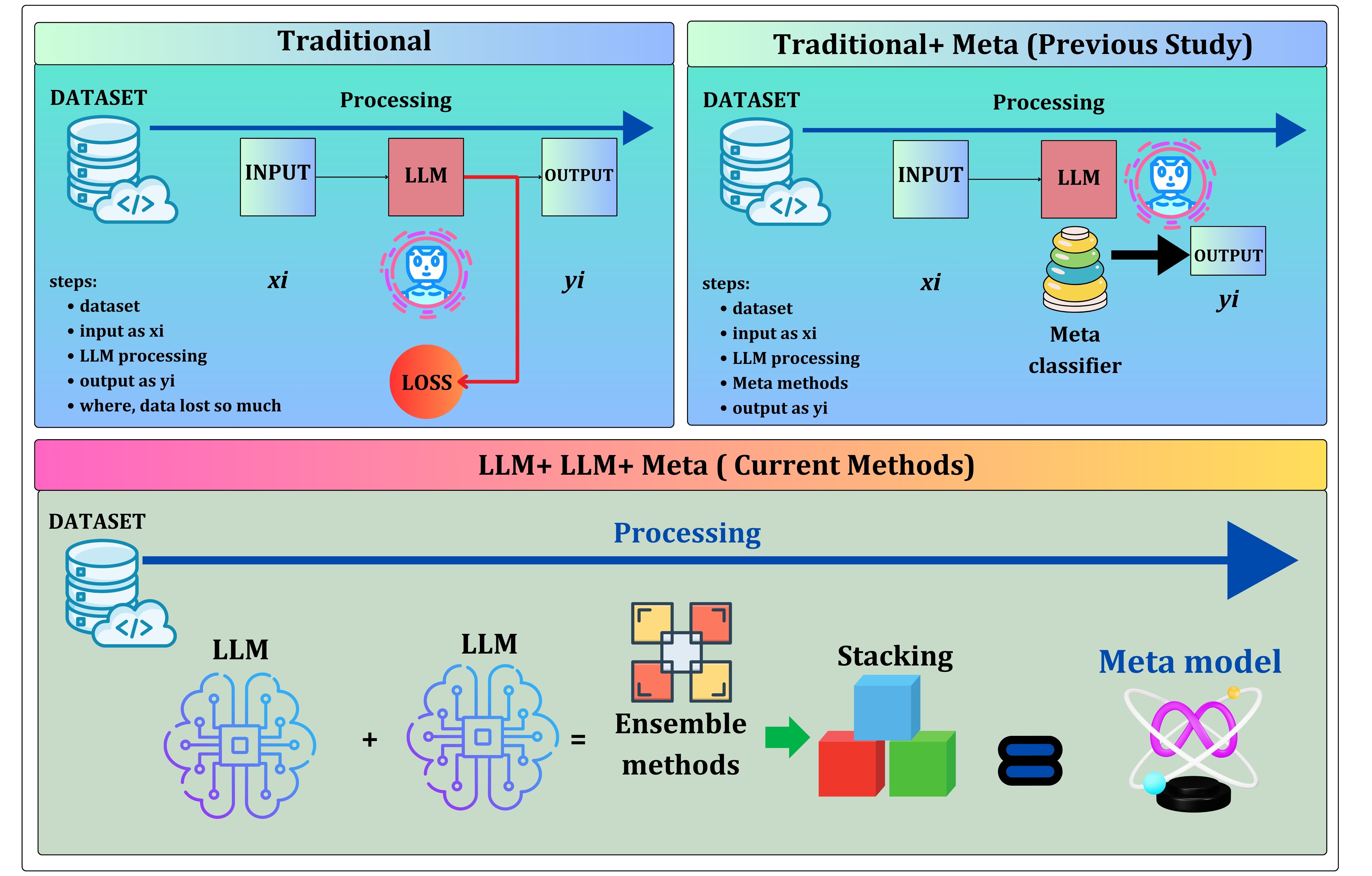}
    \vspace{-0.5cm}
    \caption{\textbf{A Comparative Overview of Vulnerability Detection Techniques.} (1) Traditional LLM-based processing, which directly outputs predictions but experiences notable data loss, (2) Traditional+Meta models from previous studies that integrate a meta-classifier to enhance LLM outputs, and (3) the proposed EnStack framework, which leverages an ensemble of multiple LLMs combined through stacking methods. EnStack incorporates a meta-model to further refine predictions, aiming for improved accuracy in vulnerability detection by effectively combining strengths of various LLMs and meta-model architectures.}
    \label{fig:f1}
\end{figure*}

Recent advancements in Artificial Intelligence (AI), particularly in Large Language Models (LLMs)\cite{balbin2020predictive,faruk2021malware,masum2021bayesian,leung2019ai,camara2018fuzzy,howlader2018predicting}, have opened new dimensions for code analysis~\cite{allamanis2018survey}. Models like CodeBERT~\cite{feng2020codebert}, GraphCodeBERT~\cite{guo2020graphcodebert}, and UniXcoder~\cite{guo2022unixcoder} have demonstrated remarkable abilities in understanding and processing programming languages by leveraging extensive code datasets. These models have been successful in tasks such as code completion, code retrieval, and bug detection by capturing both the syntactic and semantic aspects of code. Despite these achievements, challenges remain in effectively detecting complex software vulnerabilities.

Current LLM-based approaches for vulnerability detection often concentrate on specific facets of code representation. For example, CodeBERT focuses on capturing the semantic meaning~\cite{feng2020codebert} of code tokens, while GraphCodeBERT emphasizes structural relationships through data flow graphs~\cite{guo2020graphcodebert}. UniXcoder seeks to unify cross-modal representations but may not fully integrate all dimensions of code features~\cite{guo2022unixcoder}. When utilized individually, these models might not fully encapsulate the multifaceted nature of software vulnerabilities, which often require a comprehensive understanding of both syntax and context~\cite{yamaguchi2014modeling}. Additionally, earlier models like VulDeePecker~\cite{li2018vuldeepecker}, SySeVR~\cite{li2021sysevr}, and Devign~\cite{zhou2019devign} have made progress in applying deep learning to vulnerability detection but are constrained by their reliance on specific neural network architectures, such as Recurrent Neural Networks (RNNs) or Graph Neural Networks (GNNs), which may not generalize effectively across diverse vulnerability types.

To overcome these limitations, this paper introduces an ensemble-based stacking approach that combines multiple LLMs to enhance vulnerability detection in source code. By integrating CodeBERT, GraphCodeBERT, and UniXcoder through an ensemble stacking technique, we aim to leverage the unique strengths of each model to develop a more robust and comprehensive vulnerability detection system. This approach involves fine-tuning each model separately on the Draper VDISC dataset~\cite{russell2018automated}, which contains a wide range of labeled code snippets with various types of vulnerabilities. The outputs of these models are then combined using meta-classifiers, including Logistic Regression (LR)\cite{hosmer2013applied}, Support Vector Machine (SVM)\cite{pisner2020support}, Random Forest (RF)\cite{genuer2020random}, and XGBoost\cite{chen2016xgboost}.

Our method is based on the premise that integrating models that capture different aspects of code—such as semantic meaning, structural relationships, and cross-modal representations—can lead to better detection performance than using any single model alone. The ensemble stacking technique enables the meta-classifier to learn the optimal way to combine the predictions from the base models, effectively mitigating the shortcomings of individual models and enhancing the overall detection capability.

The main contributions of this paper are as follows:

\begin{itemize}
    \item \textbf{Proposing an ensemble-based stacking framework} that integrates multiple pre-trained large language models (LLMs) with meta-classifiers to enhance vulnerability detection in source code.
    \item \textbf{Conducting a comprehensive evaluation} of the proposed EnStack framework on the Draper VDISC dataset~\cite{russell2018automated}, demonstrating its superior performance compared to individual models and existing approaches across accuracy, precision, recall, F1-score, and AUC-score metrics.
    \item \textbf{Performing an ablation study} to analyze the impact of different model combinations and meta-classifiers on detection performance, providing insights into the effectiveness of the stacking approach and guiding future research on model ensemble strategies.
\end{itemize}

\section{Literature Review}

\textbf{Machine Learning Models in Vulnerability Detection.}
A vulnerability refers to a weakness in software, often stemming from errors in the code's implementation or design, that creates the potential for a security threat\cite{yuce2018fault}. For instance, accessing an array with an invalid index is a weakness, and when this weakness occurs in a specific part of a program, it becomes a vulnerability. To assist security professionals, known weaknesses are compiled in the Common Weakness Enumeration (CWE) catalog, where each weakness is given a unique ID and detailed explanation. The CVE Program (Common Vulnerabilities and Exposures) keeps a publicly available record of known vulnerabilities, accessible through platforms like the National Vulnerability Database (NVD)\cite{kekul2022comparison}.

Seas et al.\cite{seas2024automated} introduced a CNN-based model for identifying vulnerabilities in C source code. They utilized datasets from Draper Labs and the NIST SATE Juliet dataset. The source code was divided into 91 token categories and vectorized using a binary approach. The model significantly improved precision by reducing false positives while maintaining consistent recall rates.

Wu et al.\cite{wu2021vulnerability} proposed a graph-based method for classifying vulnerabilities in C/C++ functions. They used version 1.3 of the Juliet Test Suite and represented the code as simplified code property graphs to capture both syntactic and semantic features. Graph Neural Networks (GNNs) and multi-layer perceptrons were employed for learning graph representations and performing binary classification. The model achieved an F1-score of 0.836, outperforming previous CNN and RNN-based methods with F1-scores of 0.805 and 0.803.

Li et al.\cite{li2021sysevr} introduced SySeVR, a deep learning framework designed to detect vulnerabilities in C/C++ code. This approach used both syntax-based (SyVC)\cite{lin2020software} and semantics-based (SeVC)\cite{li2021vuldeelocator} vulnerability candidates, with word2vec employed for encoding SeVCs into vector representations. The dataset, containing 126 types of vulnerabilities, was split into 80\% training and 20\% testing. SySeVR achieved an F1-score of 0.926, significantly surpassing VulDeePecker\cite{li2018vuldeepecker}, which scored 0.783. The use of stratified 5-fold cross-validation ensured the robustness of the results.

\textbf{Large Language Models in Vulnerability Detection.}
Kim et al.\cite{kim2022vuldebert} developed VulDeBERT, a BERT-based model fine-tuned on C/C++ source code for detecting specific vulnerabilities. They processed the source code using a method called code gadget. Tested on datasets representing CWE-119 (buffer errors) and CWE-399 (management errors), achieved results with F1-scores of 0.946 and 0.979, outperforming VulDeePecker by a large margin.

Hanif and Maffeis\cite{hanif2022vulberta} presented VulBERTa, a vulnerability detection model that uses RoBERTa and a novel tokenization method. The model was evaluated using two architectures—VulBERTa-MLP and VulBERTa-CNN—achieving F1-scores of 93.03 and 90.86, respectively, on the VulDeePecker dataset and 43.34 and 57.92 on the Draper dataset.

Fu et al.\cite{fu2023chatgpt} examined the performance of GPT-3.5 and GPT-4 for vulnerability classification tasks. The GPT models underperformed compared to fine-tuned models like CodeBERT, achieving lower accuracies in both function-level and line-level vulnerability classification tasks.

\section{Methodology}
Detecting vulnerabilities in source code is crucial for ensuring software security. Various models have been proposed to address this challenge, but improvements are needed to handle the complex structure of source code effectively. In this study, we employ an ensemble stacking approach that combines the strengths of multiple large language models (LLMs) to improve vulnerability detection. The models used include CodeBERT, GraphCodeBERT, and UniXcoder, and their outputs are combined using an ensemble stacking technique.

\subsection{Problem Formulation}

Given a dataset \( D = \{(x_i, y_i)\}_{i=1}^{n} \), where each \( x_i \) represents a code snippet and \( y_i \in \{0, 1, 2, 3, 4\} \) denotes the vulnerability class label (e.g., CWE categories), the objective is to develop a model that accurately predicts the vulnerability class \( \hat{y}_i \) for unseen code snippets.

To achieve this, we employ the EnStack ensemble stacking framework. The framework utilizes a set of base models \( \mathcal{M} = \{ M_k \}_{k=1}^{K} \) (e.g., CodeBERT, GraphCodeBERT, UniXcoder), each fine-tuned on the training data \( D_{\text{train}} \) to detect vulnerabilities independently. Each base model \( M_k \) produces an output vector for a given code snippet \( x_i \), typically a probability distribution over the vulnerability classes.

We construct a meta-feature vector \( \mathbf{z}_i \) for each sample \( x_i \) by concatenating the outputs of all base models:

\[
\mathbf{z}_i = [ M_1(x_i), M_2(x_i), \dots, M_K(x_i) ].
\]

A meta-classifier \( \mathcal{F}_{\text{meta}} \) is then trained on these meta-features to learn the optimal combination of base model predictions:

\[
\hat{y}_i = \mathcal{F}_{\text{meta}}( \mathbf{z}_i ).
\]

The goal is to find the meta-classifier \( \mathcal{F}_{\text{meta}} \) that minimizes the classification error on the validation set \( D_{\text{val}} \) and performs well on the test set \( D_{\text{test}} \). This involves selecting \( \mathcal{F}_{\text{meta}} \) from a set of candidate meta-classifiers \( \mathcal{C} = \{ C_m \}_{m=1}^{M} \) (e.g., logistic regression, random forest).


\begin{figure*}
    \centering
    \includegraphics[width=1\linewidth]{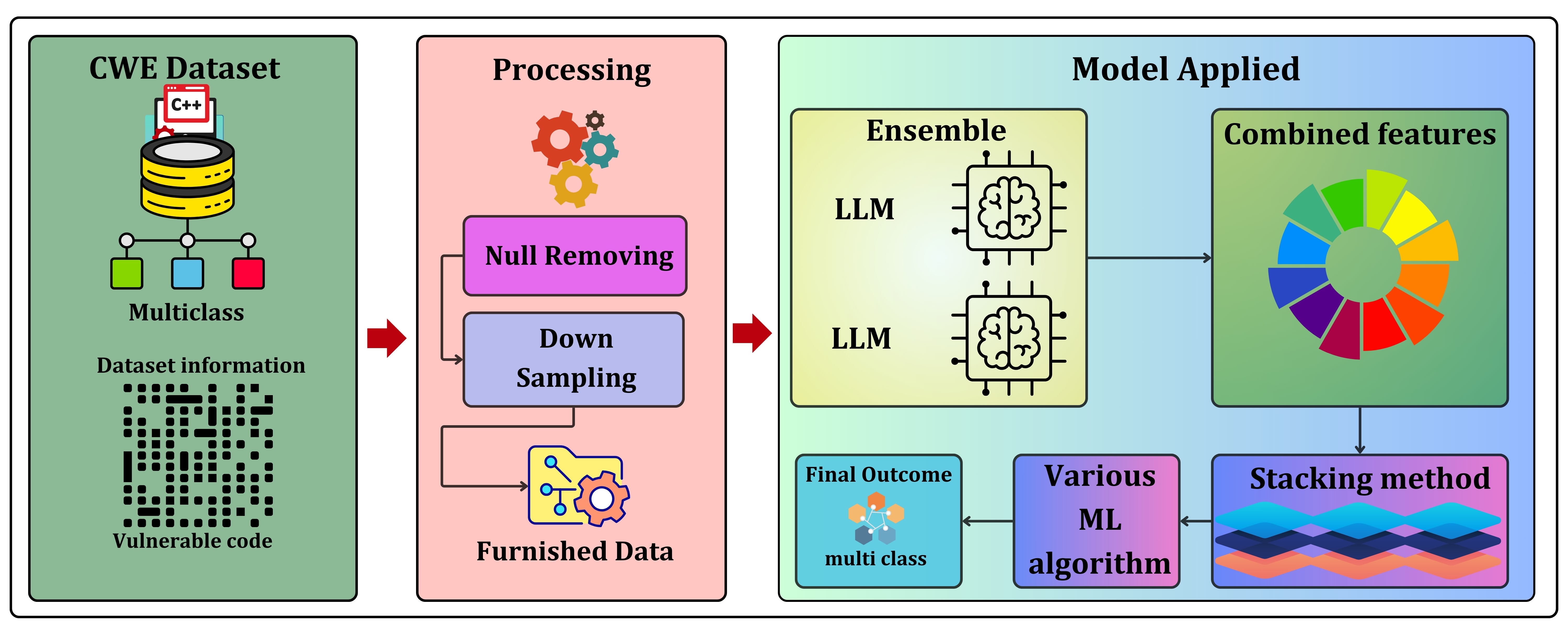}
    \vspace{-0.5cm}
    \caption{\textbf{Methodology Workflow for Enhanced Vulnerability Detection using EnStack Framework.} The proposed methodology begins with the CWE dataset, which includes multiclass labels associated with various types of vulnerable code. During the processing phase, data cleansing steps, such as null value removal and down sampling, ensure balanced and representative input data. This furnished data is then fed into an ensemble of large language models (LLMs), where features are extracted and combined to capture diverse vulnerability patterns. These combined features are passed through a stacking framework that employs various machine learning algorithms to enhance model robustness. The stacked model produces a final multiclass prediction, facilitating accurate and refined detection of code vulnerabilities across different classes.}
    \label{fig:methodology}
\end{figure*}

\subsection{Dataset and Preprocessing}
\textbf{Dataset Details.}\\
The Draper VDISC dataset is a large-scale collection of over 1.27 million code functions sourced from open-source software projects. These code functions are labeled based on potential vulnerabilities detected through static analysis. The dataset contains five primary vulnerability classes, each mapped to Common Weakness Enumeration (CWE) categories:
\begin{itemize}
    \item CWE-119: Memory-related vulnerabilities
    \item CWE-120: Buffer overflow vulnerabilities
    \item CWE-469: Integer overflow vulnerabilities
    \item CWE-476: NULL pointer dereference vulnerabilities
    \item CWE-other: Miscellaneous vulnerabilities
\end{itemize}

For this study, the dataset is divided into training (80\%), validation (10\%), and test (10\%) sets. The class distribution is imbalanced, with certain vulnerabilities, such as CWE-469 (integer overflow), being significantly less frequent than others like CWE-119 (memory vulnerabilities). Null entries were then removed from the training, validation, and test sets.

\textbf{Data Balancing.}\\
To address the class imbalance, we employed downsampling to reduce the size of the majority classes, such as CWE-119 and CWE-120, bringing them closer to the size of the minority classes, particularly CWE-469. Downsampling was selected as it effectively reduces the dominance of the majority classes without introducing synthetic data, which could potentially introduce noise or lead to overfitting. Additionally, downsampling is computationally efficient, allowing faster model training while maintaining a representative sample of the different vulnerabilities.

Other techniques, such as oversampling or synthetic data generation were considered but ultimately not used due to the large dataset size. Downsampling was the preferred method as it prevents overfitting and preserves the dataset's natural characteristics. Table \ref{dataset_distribution} presents the overall distribution of the dataset across the training, test, and validation sets.

\begin{table}[htbp]
\caption{Class Distribution Across Training, Validation, and Test Sets}
\centering
\resizebox{\columnwidth}{!}{%
\begin{tabular}{c c c c c}
\hline
\textbf{Class Label} & \textbf{CWE Type} & \textbf{Training Samples} & \textbf{Validation Samples} & \textbf{Test Samples} \\
\hline
0 & CWE-119 (Memory) & 5942 & 1142 & 1142 \\
1 & CWE-120 (Buffer Overflow) & 5777 & 1099 & 1099 \\
2 & CWE-469 (Integer Overflow) & 249 & 53 & 53 \\
3 & CWE-476 (Null Pointer) & 2755 & 535 & 535 \\
4 & CWE-other & 5582 & 1071 & 1071 \\
\hline
\textbf{Total} & & 20305 & 3900 & 3900 \\
\hline
\end{tabular}%
}
\label{dataset_distribution}
\end{table}

\subsection{Fine-Tuning and Training}
To enhance the vulnerability detection capabilities of large language models (LLMs) for code-related tasks, we fine-tuned three pre-trained models: CodeBERT, GraphCodeBERT, and UniXcoder. These models were selected for their specialized architectures, which are well-suited to understanding code syntax, semantics, and structure. The fine-tuning process comprised two stages: (a) fine-tuning the models using vulnerability class labels and (b) combining the outputs of multiple LLMs through an ensemble stacking method to further improve classification performance.
\textbf{Models Used:}
\begin{enumerate}
    \item \textbf{CodeBERT~\cite{feng2020codebert}}, a model pre-trained on code and natural language, was fine-tuned using the Draper VDISC dataset. The fine-tuning process optimized the model for multi-class classification using \textit{cross-entropy loss}.
    \item \textbf{GraphCodeBERT~\cite{guo2020graphcodebert}} is designed to handle the graph-based structure of code. It was fine-tuned on the same dataset to enhance its ability to detect code vulnerabilities.
    \item \textbf{UniXcoder~\cite{guo2022unixcoder}} was fine-tuned to improve its understanding of the syntactic and semantic structures of code snippets, making it suitable for vulnerability detection.
\end{enumerate}

\textbf{Hyperparameter Setup.}\\
The training setup for each model is summarized in Table \ref{tab2}.

\begin{table}[htbp]
\caption{Hyperparameter Setup for Fine-Tuning LLMs and Machine Learning Models}
\centering
\small 
\begin{tabular}{c c}
\hline
\textbf{Hyperparameter} & \textbf{Value} \\
\hline
\textbf{Transformer Model} & \\
Batch Size & 16 \\
Epochs & 10 \\
Learning Rate & 2 $\times$ 10$^{-5}$ \\
Optimizer & AdamW \\
Max Token Length & 512 \\
\hline
\textbf{Logistic Regression} & \\
Max Iterations & 200 \\
Solver & liblinear \\
\hline
\textbf{Random Forest} & \\
Number of Estimators & 200 \\
Max Depth & 10 \\
\hline
\textbf{SVM} & \\
Kernel & RBF \\
Probability Estimation & True \\
Random State & 42 \\
\hline
\textbf{XGBoost} & \\
Number of Estimators & 100 \\
Eval Metric & mlogloss \\
Learning Rate & 0.1 \\
Max Depth & 6 \\
\hline
\end{tabular}
\label{tab2}
\end{table}

\textbf{Fine-Tuning using Labels.}

For each model, the fine-tuning process involved training the models on the downsampled Draper VDISC dataset using the Cross Entropy Loss function. The input code snippets were tokenized and fed into the model to predict their vulnerability classes. The original labels from the dataset, denoting vulnerability types (CWE categories), were used to guide the training process, minimizing the cross-entropy loss between the predicted outputs $\hat{y}_i$ and the ground truth labels $y_i$.

\subsection{Ensemble Stacking Method}

To further enhance the detection performance, we implemented an ensemble stacking method to combine the outputs of CodeBERT, GraphCodeBERT, and UniXcoder. This approach allowed us to leverage the strengths of each model to improve vulnerability classification.

\textbf{Model Predictions.}

Each model generated a probability distribution over the five vulnerability classes. These distributions, represented as \( M_k(x_i) \), were concatenated to form feature vectors. For each code snippet $x_i$, the model predictions from CodeBERT, GraphCodeBERT, and UniXcoder were combined into a single feature vector.

\textbf{Meta-Classifier Training.}

Four meta-classifiers were tested to combine the output vectors:
\begin{itemize}
    \item Logistic Regression (LR)\cite{hosmer2013applied}
    \item Random Forest (RF)\cite{genuer2020random}
    \item Support Vector Machine (SVM)\cite{pisner2020support}
    \item XGBoost\cite{chen2016xgboost}
\end{itemize}
These classifiers were chosen for their ability to handle high-dimensional data and combine diverse outputs from base models. The meta-classifiers were trained on the concatenated output features and were evaluated on the test set.

\subsection{Algorithm}
The EnStack framework, outlined in Algorithm \ref{alg:enstack}, is an ensemble stacking approach developed to enhance vulnerability detection accuracy. The process begins by preparing the dataset, where data is split into training, validation, and test subsets, and the training set is balanced through downsampling. Each base model within the ensemble is then fine-tuned on the training data. Meta-features are generated by aggregating predictions from all base models for each data sample, creating a comprehensive meta-feature vector. Multiple meta-classifiers are trained on these meta-features, with the classifier demonstrating the best validation performance selected as the optimal meta-classifier. This selected meta-classifier is finally evaluated on the test set to assess its performance, providing a robust solution for effective vulnerability detection.

\begin{algorithm}
\caption{EnStack: Ensemble Stacking Framework for Vulnerability Detection}
\label{alg:enstack}
\begin{algorithmic}[1]
\REQUIRE Dataset $D = \{(x_i, y_i)\}_{i=1}^{n}$; Base models $\mathcal{M} = \{M_k\}_{k=1}^{K}$; Meta-classifiers $\mathcal{C} = \{C_m\}_{m=1}^{M}$
\ENSURE Optimal meta-classifier $\mathcal{F}_{\text{meta}}$ and performance metrics
\STATE \textbf{Data Preparation}:
    \STATE Split $D$ into $D_{\text{train}}$, $D_{\text{val}}$, and $D_{\text{test}}$
    \STATE Balance $D_{\text{train}}$ via downsampling
\STATE \textbf{Base Model Training}:
    \FOR{each base model $M_k \in \mathcal{M}$}
        \STATE Fine-tune $M_k$ on $D_{\text{train}}$
    \ENDFOR
\STATE \textbf{Meta-feature Generation}:
    \FOR{each sample $(x_j, y_j) \in D_{\text{val}}$}
        \STATE Compute meta-feature vector:
        \STATE \quad $\mathbf{z}_j = [ M_1(x_j), M_2(x_j), \dots, M_K(x_j) ]$
    \ENDFOR
\STATE \textbf{Meta-classifier Training and Selection}:
    \FOR{each meta-classifier $C_m \in \mathcal{C}$}
        \STATE Train $C_m$ on $\{ (\mathbf{z}_j, y_j) \}_{j}$
        \STATE Evaluate $C_m$ on $D_{\text{val}}$
    \ENDFOR
    \STATE Select optimal $\mathcal{F}_{\text{meta}}$ with best validation performance
\STATE \textbf{Evaluation on Test Set}:
    \FOR{each sample $(x_i, y_i) \in D_{\text{test}}$}
        \STATE Compute $\mathbf{z}_i = [ M_1(x_i), M_2(x_i), \dots, M_K(x_i) ]$
        \STATE Predict $\hat{y}_i = \mathcal{F}_{\text{meta}}( \mathbf{z}_i )$
    \ENDFOR
    \STATE Compute performance metrics on $D_{\text{test}}$
\RETURN Optimal meta-classifier $\mathcal{F}_{\text{meta}}$ and performance metrics
\end{algorithmic}
\end{algorithm}

\section{Experiment}

\subsection{Hardware and Software Setup}
All experiments were conducted using the NVIDIA Tesla P100 GPU. We used the PyTorch deep learning framework for model implementation and Hugging Face’s transformers library to fine-tune pre-trained models such as CodeBERT, GraphCodeBERT, and UniXcoder. For the ensemble stacking technique, scikit-learn was employed to implement meta-classifiers, including Logistic Regression, Random Forest, Support Vector Machine (SVM), and XGBoost.

\begin{figure*}
    \centering
    \includegraphics[width=1\linewidth]{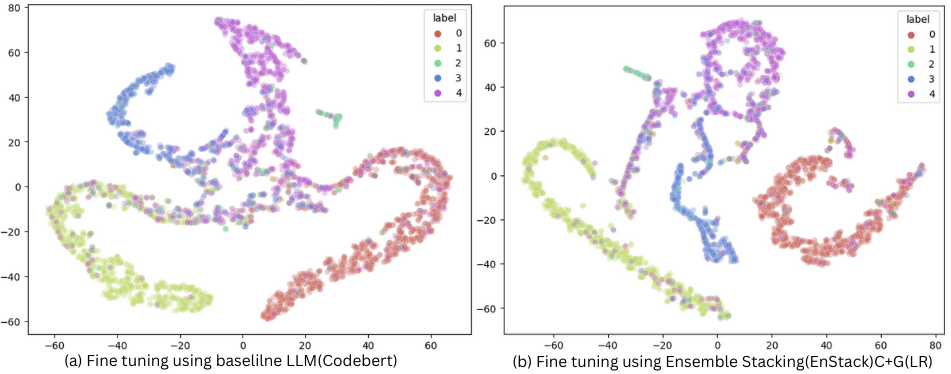}
    \vspace{-0.5cm}
    \caption{\textbf{t-SNE visualization of latent space representations for CWE categories in vulnerability detection.} (a) baseline model fine-tuned using CodeBert, exhibiting class overlap and reduced inter-cluster separation; (b) enhanced representations through ensemble stacking (EnStack) with CodeBert and  GraphCodeBERT (C+G) followed by logistic regression (LR), demonstrating improved cluster formation and class separability.}
    \label{fig:tsne}
\end{figure*}

\subsection{Baselines}
We compare our model with several baselines, including transformer-based models such as CodeBERT, GraphCodeBERT, and UniXcoder. Additionally, we included Attention LSTM\cite{wang2016attention} as a non-transformer baseline model to highlight the performance benefits of pre-trained language models.

\textbf{Evaluation Metrics.}
The models' performance was evaluated using several metrics, including:
\begin{itemize}
    \item Accuracy: Proportion of correct predictions.
    \item Precision, Recall, and F1-Score: Precision represents the accuracy of positive predictions, recall measures the coverage of actual positives, and the F1-Score balances both precision and recall.
    \item AUC-Score: The Area Under the ROC Curve (AUC) was computed to assess the models' ability to discriminate between vulnerability classes.
\end{itemize}

\begin{table*}[htbp]
\centering
\caption{Comprehensive evaluation of individual models, single stacked models, and ensemble stacking methods across key performance metrics: \textbf{Accuracy}, \textbf{Precision}, \textbf{Recall}, \textbf{F1-Score}, and \textbf{AUC-Score}. All results are reported as percentages, with $\uparrow$ indicating higher values are better.}
\renewcommand{\arraystretch}{1.4} 
\setlength{\tabcolsep}{8pt} 
\resizebox{\textwidth}{!}{%
\begin{tabular}{lccccc}
\hline
\textbf{Model} & \textbf{Accuracy (\%) $\uparrow$} & \textbf{Precision (\%) $\uparrow$} & \textbf{Recall (\%) $\uparrow$} & \textbf{F1-Score (\%) $\uparrow$} & \textbf{AUC-Score (\%) $\uparrow$} \\
\hline
\multicolumn{6}{c}{\textbf{Individual Models}} \\
Attention LSTM & 73.00 & 72.97 & 73.00 & 72.95 & 77.54 \\
CodeBERT (C) & 78.51 & 77.85 & 78.51 & 77.98 & 92.16 \\
GraphCodeBERT (G) & 80.05 & 79.92 & 80.05 & 79.86 & 93.36 \\
UniXcoder (U) & 81.54 & 81.96 & 81.54 & 81.49 & \textbf{93.80} \\
\hline
\multicolumn{6}{c}{\textbf{Stacked Models with Individual Base Models}} \\
Stacking C (LR) & 79.00 & 78.71 & 79.00 & 78.62 & 91.26 \\
Stacking C (RF) & 78.69 & 79.10 & 78.69 & 78.70 & 91.64 \\
Stacking C (SVM) & 78.74 & 79.85 & 78.74 & 78.64 & 88.55 \\
Stacking C (XGBoost) & 77.26 & 77.14 & 77.26 & 77.12 & 90.99 \\
Stacking G (LR) & 80.13 & 80.12 & 80.13 & 79.88 & 92.39 \\
Stacking G (RF) & 79.79 & 80.01 & 79.79 & 79.63 & 92.11 \\
Stacking G (SVM) & 80.00 & 80.82 & 80.00 & 80.04 & 89.60 \\
Stacking G (XGBoost) & 78.46 & 78.47 & 78.46 & 78.36 & 90.36 \\
Stacking U (LR) & 81.51 & 81.70 & 81.51 & 81.36 & 92.49 \\
Stacking U (RF) & 81.44 & 81.71 & 81.44 & 81.35 & 92.42 \\
Stacking U (SVM) & 81.31 & 81.89 & 81.31 & 81.33 & 90.76 \\
Stacking U (XGBoost) & 80.08 & 79.90 & 80.08 & 79.94 & 92.57 \\
\hline
\multicolumn{6}{c}{\textbf{Ensemble Stacking Models}} \\
Ensemble Stacking C+G (LR) & 81.13 & 81.10 & 81.13 & 80.90 & 92.93 \\
Ensemble Stacking C+G (RF) & 81.56 & 81.87 & 81.56 & 81.44 & 92.32 \\
Ensemble Stacking C+G (SVM) & 81.46 & 81.77 & 81.46 & 81.40 & 89.96 \\
Ensemble Stacking C+G (XGBoost) & 80.28 & 80.22 & 80.28 & 80.06 & 91.31 \\
Ensemble Stacking G+U (LR) & \textbf{82.36} & 82.59 & \textbf{82.36} & 82.21 & 92.85 \\
Ensemble Stacking G+U (RF) & 82.28 & 82.45 & 82.28 & 82.13 & 92.45 \\
Ensemble Stacking G+U (SVM) & \textbf{82.36} & \textbf{82.85} & \textbf{82.36} & \textbf{82.28} & 90.53 \\
Ensemble Stacking G+U (XGBoost) & 80.67 & 80.47 & 80.67 & 80.46 & 92.28 \\
\hline
\end{tabular}%
}
\label{tab:individual_and_stacked_results}
\end{table*}

\subsection{Results and Analysis}

\textbf{Individual Model Performance.}

We first evaluated the individual models, including Attention LSTM, CodeBERT (C), GraphCodeBERT (G), and UniXcoder (U), using accuracy, precision, recall, F1-score, and AUC-score. These results, summarized in Table \ref{tab:individual_and_stacked_results}, highlight the distinct strengths and limitations of each model in detecting vulnerabilities within source code.

The non-transformer baseline, Attention LSTM, achieved an accuracy of 73.00\%, reflecting its ability to capture sequential dependencies but also its limitations in handling the more intricate semantic and structural aspects of code. The model’s low AUC-score of 77.54\% further underscores its struggles with distinguishing between vulnerability classes effectively.

In contrast, CodeBERT, a transformer-based model, demonstrated significant improvement with an accuracy of 78.51\% and an F1-score of 77.98\%. Its ability to process both natural language and programming code semantics contributes to these gains, though its relatively lower recall (78.51\%) indicates that it might not fully capture structural nuances essential for vulnerability detection.

GraphCodeBERT (G), which integrates both code semantics and structural information, outperformed CodeBERT with an accuracy of 80.05\%, an F1-score of 79.86\%, and a higher AUC-score of 93.36\%. The model’s enhanced understanding of code flow and data dependencies makes it more robust for identifying vulnerabilities.

UniXcoder (U) achieved the highest performance among the individual models, with an accuracy of 81.54\% and an F1-score of 81.49\%. UniXcoder’s ability to represent both the syntactic and semantic properties of code through cross-modal learning allowed it to surpass the other models. Its AUC-score of 93.80\%, the highest among all models, emphasizes its superior capacity to distinguish between different classes of vulnerabilities.

\textbf{Stacking with a Single LLM and Meta-Classifier.}
We explored the impact of stacking a single LLM with traditional machine learning meta-classifiers, such as Logistic Regression (LR), Support Vector Machine (SVM), Random Forest (RF), and XGBoost. The results demonstrate that stacking improves performance over individual models. For example, stacking UniXcoder with SVM resulted in an accuracy of 81.36\% and an F1-score of 81.89\%, compared to the standalone UniXcoder model. Similarly, stacking GraphCodeBERT with Logistic Regression achieved an accuracy of 80.13\%, further illustrating the benefits of combining an LLM with a meta-classifier.

These findings suggest that even without leveraging multiple LLMs, a single LLM can be effectively enhanced by traditional classifiers, refining its predictions and increasing its discriminatory ability. Among the meta-classifiers, SVM consistently delivered strong performance due to its capability of handling high-dimensional feature spaces.

\textbf{Ensemble Stacking of Multiple LLMs with Meta Classifiers.}

The most notable performance improvements, however, were observed when combining multiple LLMs in an ensemble stacking approach. Stacking GraphCodeBERT and UniXcoder (G+U) with SVM delivered the highest accuracy of 82.36\% and the best F1-score of 82.28\%. This result underscores the effectiveness of combining the structural insights from GraphCodeBERT with the cross-modal learning of UniXcoder, allowing the meta-classifier to leverage both models' strengths. The highest AUC-score of 92.85\% was achieved by stacking G+U with Logistic Regression, indicating the model's superior discriminatory power across classes.

\subsection{Ablation Study}
\textbf{Impact of Model Combinations in Stacking.}
We conducted a detailed analysis of various model combinations in our stacking ensembles, focusing on CodeBERT (C), GraphCodeBERT (G), and UniXcoder (U). As illustrated in Figure \ref{fig:ablation_study} (a), stacking GraphCodeBERT and UniXcoder (Stacking G+U) consistently achieved superior performance compared to other combinations across different meta-classifiers. Specifically, stacking G+U with SVM attained the highest accuracy of 82.36\% and an F1-score of 82.28\%, significantly outperforming stacking combinations like C+G, which reached a lower accuracy of 81.46\% with SVM. This demonstrates that combining GraphCodeBERT’s structural insights with UniXcoder’s syntactic-semantic understanding results in a more comprehensive feature set, optimizing the performance of stacking ensembles for vulnerability detection. These results emphasize the importance of selecting complementary base models to maximize stacking effectiveness.

\textbf{Impact of Meta-Classifier Choice in Stacking.}
In addition to model combinations, we investigated the impact of different meta-classifiers, including Logistic Regression (LR), Random Forest (RF), Support Vector Machine (SVM), and XGBoost, on the performance of stacking ensembles. As shown in Figure \ref{fig:ablation_study} (b), the choice of meta-classifier significantly affects performance. Stacking G+U with SVM yielded the highest F1-score 82.28\%) and precision (82.85\%), while Logistic Regression achieved the highest AUC-Score (92.85\%) in the same configuration. In contrast, XGBoost consistently underperformed, with an accuracy of only 80.67\% when stacking G+U. This suggests that linear classifiers such as SVM and LR are better suited for leveraging the complementary features of base models in this task. More complex classifiers like XGBoost may introduce unnecessary complexity, which does not translate into better performance, further validating the suitability of simpler, interpretable classifiers for this application.

\begin{figure}[ht]
    \centering
    \includegraphics[width=\linewidth]{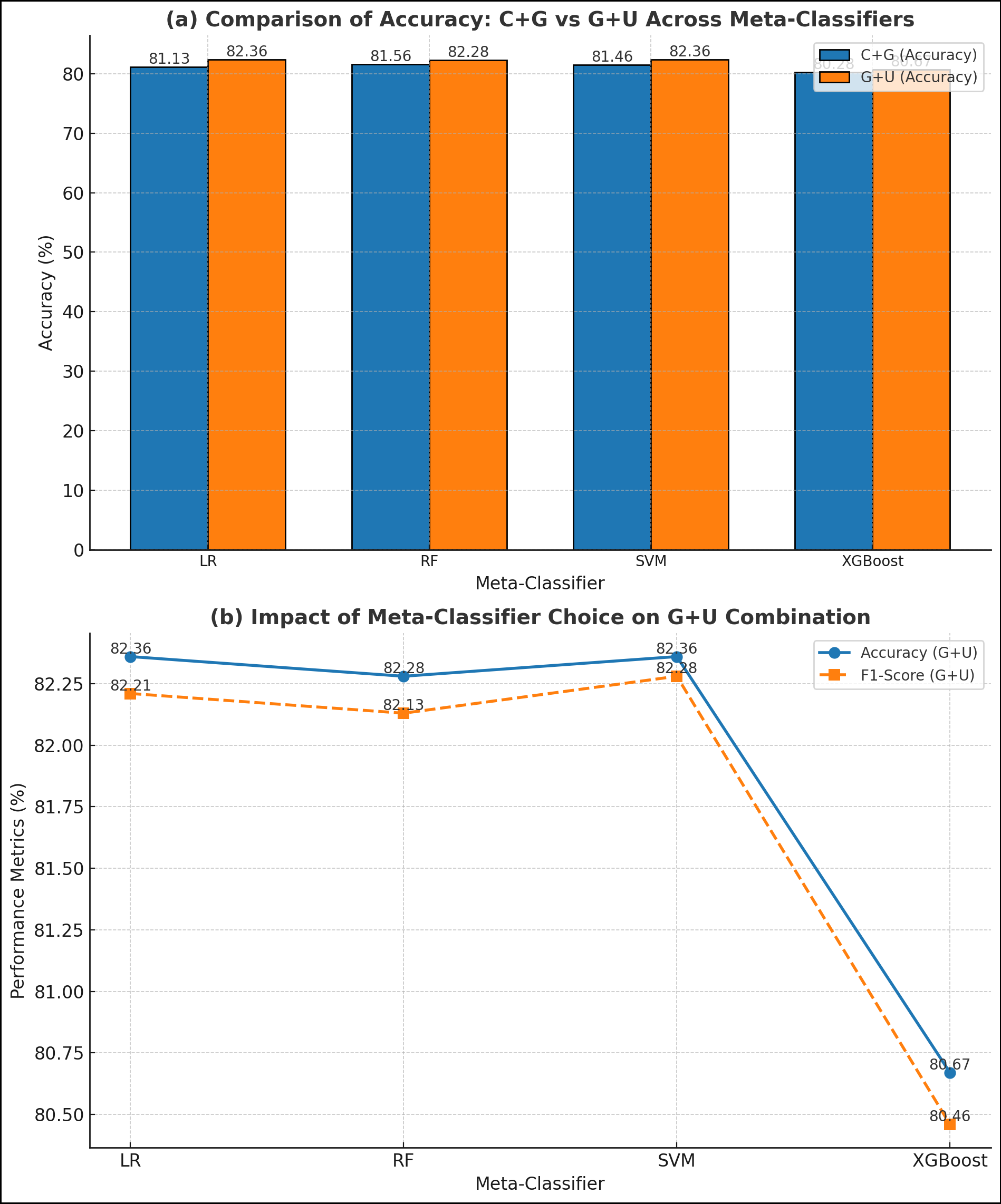}
    \caption{\textbf{Ablation study results showcasing the impact of model combinations and meta-classifier choice in stacking ensembles.}
    (a) Performance comparison of model combinations (C+G vs. G+U) across meta-classifiers (LR, RF, SVM, and XGBoost). \textbf{G+U consistently outperforms C+G}, with SVM achieving the highest accuracy of \textbf{82.36\%}.
    (b) Detailed analysis of meta-classifier performance on the G+U combination, highlighting the dominance of linear classifiers (SVM and LR) in capturing complementary features. \textbf{XGBoost underperforms}, indicating added complexity may not yield better results for this task.}
    \label{fig:ablation_study}
\end{figure}

\subsection{Discussion}
\textbf{Ensemble Stacking for Vulnerability Detection.} Our experimental results demonstrate that the ensemble stacking approach, which combines the strengths of multiple transformer-based models, outperforms individual models in detecting vulnerabilities in source code. The individual models, such as CodeBERT, GraphCodeBERT, and UniXcoder, achieved respectable accuracy and F1-scores, with UniXcoder performing the best at 81.54\% accuracy. However, combining these models through both weighted aggregation and stacking significantly enhanced the overall performance. The ensemble methods—particularly Stacking G+U with Logistic Regression (LR) and Support Vector Machine (SVM)—achieved the highest accuracy of 82.36\%, showcasing the value of integrating diverse model strengths. SVM further excelled by attaining the highest F1-score of 82.28\%, while Logistic Regression demonstrated superior discriminative power with an AUC score of 92.85\%. These results highlight the effectiveness of ensemble stacking in improving vulnerability detection, especially by leveraging complementary features such as syntactic-semantic cross-modal understanding from UniXcoder and structural insights from GraphCodeBERT.

\textbf{Limitations.} The EnStack framework has several limitations. First, the dataset suffers from severe class imbalance, with vulnerabilities like CWE-469 (Integer Overflow) being significantly underrepresented, reducing the model's ability to generalize effectively. Downsampling used to mitigate this imbalance further reduced the dataset size, compounding generalization challenges. Second, reliance on the Draper VDISC dataset restricts applicability, as it focuses on specific vulnerabilities and programming languages, raising concerns about generalization to other datasets and languages that may exhibit distinct vulnerability patterns. Third, the ensemble stacking approach imposes substantial computational overhead, with the integration of pre-trained models (CodeBERT, GraphCodeBERT, UniXcoder) and meta-classifier training requiring significant resources, limiting scalability for real-time or large-scale applications.

\section{Conclusion}
In this study, we proposed an ensemble stacking approach that combines the outputs of transformer-based models like CodeBERT, GraphCodeBERT, and UniXcoder with meta classifiers to improve vulnerability detection in source code. By fine-tuning each model on the Draper VDISC dataset and utilizing both ensemble and stacking methods, we demonstrated that the ensemble stacking approach significantly enhances the performance of individual models. Among the meta-classifiers, Logistic Regression (LR) and Support Vector Machine (SVM) produced the best results, achieving an accuracy of 82.36\% and a high AUC score of 92.85\%. These results validate the effectiveness of leveraging diverse models that capture different aspects of code—syntactic, semantic, and structural—to improve classification of vulnerabilities in a multi-class setting.

\textbf{Future Work.}
While our EnStack framework demonstrated promising results, there is still room for improvement in addressing the challenges posed by underrepresented classes in vulnerability detection. In future work, we plan to experiment with multiple vulnerability datasets to evaluate the generalizability and robustness of our models across diverse codebases. By incorporating a wider range of datasets, we aim to capture a broader spectrum of vulnerability types and patterns, enhancing the models' ability to detect rare and complex issues.

Furthermore, we intend to explore generative models such as LLaMA and Mistral to enhance code understanding and vulnerability detection. These models, with their powerful generative capabilities, can be leveraged for domain-specific pre-training, allowing for more comprehensive detection of nuanced vulnerabilities. Additionally, we aim to investigate how generative models can provide richer contextual information about code, potentially generating synthetic code snippets to augment training data and improve model robustness. Exploring transfer learning with these generative models will further allow us to tailor detection strategies, leading to more accurate and nuanced vulnerability identification.
\section*{Acknowledgment}
This work was partially supported by project SERICS (PE00000014) under the MUR National Recovery and Resilience Plan funded by the European Union - NextGenerationEU


\bibliographystyle{ieeetr}
\bibliography{references}


\end{document}